\documentclass[%
aip,
amsmath,amssymb,
reprint,%
]{revtex4-1}

\usepackage{graphicx}               
\usepackage{dcolumn}                
\usepackage{bm}                     

\usepackage[utf8]{inputenc}
\usepackage[T1]{fontenc}
\usepackage{amsmath}
\usepackage{amssymb}
\usepackage{mathptmx}
\usepackage{braket}
\usepackage{longtable}
\usepackage{siunitx}
\usepackage{etoolbox}
\usepackage{gensymb}

\makeatletter
\def\@email#1#2{%
 \endgroup
 \patchcmd{\titleblock@produce}
  {\frontmatter@RRAPformat}
  {\frontmatter@RRAPformat{\produce@RRAP{*#1\href{mailto:#2}{#2}}}\frontmatter@RRAPformat}
  {}{}
}%
\makeatother

\DeclareSIUnit[number-unit-product = {\,}]\cal{cal}
\DeclareSIUnit[number-unit-product = {\,}]\mol{mol}

\usepackage{xcolor}
\usepackage[normalem]{ulem}

\newcommand{\trial}{\mathrm{T}}
\newcommand{\init}{\mathrm{I}}

\begin{document}

\preprint{AIP/123-QED}

\title[]{Benchmark Phaseless Auxiliary-Field Quantum Monte Carlo Method for Small Molecules}

\author{Zoran Sukurma}
\affiliation{University of Vienna, Faculty of Physics and Center for Computational Materials Science, Kolingasse 14-16, A-1090 Vienna, Austria}
\affiliation{University of Vienna, Faculty of Physics \& Vienna Doctoral School in Physics,  Boltzmanngasse 5, A-1090 Vienna, Austria}
\author{Martin Schlipf}
\affiliation{VASP Software GmbH, Sensengasse 8, 1090 Vienna, Austria}
\author{Moritz Humer}
\affiliation{University of Vienna, Faculty of Physics and Center for Computational Materials Science, Kolingasse 14-16, A-1090 Vienna, Austria}
\affiliation{University of Vienna, Faculty of Physics \& Vienna Doctoral School in Physics,  Boltzmanngasse 5, A-1090 Vienna, Austria}
\author{Amir Taheridehkordi}
\affiliation{University of Vienna, Faculty of Physics and Center for Computational Materials Science, Kolingasse 14-16, A-1090 Vienna, Austria}
\author{Georg Kresse}
\affiliation{University of Vienna, Faculty of Physics and Center for Computational Materials Science, Kolingasse 14-16, A-1090 Vienna, Austria}
\affiliation{VASP Software GmbH, Sensengasse 8, 1090 Vienna, Austria}

\date{\today}

\begin{abstract}
We report a scalable Fortran implementation of the phaseless auxiliary-field quantum Monte Carlo (ph-AFQMC) and demonstrate its excellent performance and beneficial scaling with respect to system size. 
Furthermore, we investigate modifications of the phaseless approximation that can help to reduce the overcorrelation problems common to the ph-AFQMC.
We apply the method to the 26 molecules in the HEAT set, the benzene molecule, and water clusters.
We observe a mean absolute deviation of the total energy of \SI{1.15}{\kilo \cal / \mol } for the molecules in the HEAT set; close to chemical accuracy.
For the benzene molecule, the modified algorithm despite using a single-Slater-determinant trial wavefunction yields the same accuracy as the original phaseless scheme with 400 Slater determinants.
Despite these improvements, we find systematic errors for the CN, CO$_2$, and O$_2$ molecules that need to be addressed with more accurate trial wavefunctions.
For water clusters, we find that the ph-AFQMC yields excellent binding energies that differ from CCSD(T) by typically less than \SI{0.5}{\kilo \cal / \mol}.

\end{abstract}

\maketitle

\section{Introduction}
\label{intro}
Density functional theory (DFT) approximately solves the many-body Schr\"odinger equation by mapping it to a single-particle problem and has become routine in quantum chemistry and condensed matter physics.\cite{KohnSham1965,Kresse1996Vasp,Kresse1999PAW}
Despite its simplicity and the fact that it can describe systems with hundreds of atoms, DFT is not always accurate enough to serve as a general-purpose solution.
The stretched H$_2$ molecule is a well-known example of the limitations of DFT, as its results are qualitatively wrong. \cite{Burke2015Bonds}
The key problem is the treatment of the electron-electron correlation effects via approximate exchange-correlation (xc) energy functionals.\cite{Cohen2008DFTLimits} 
Perdew established the analogy of \emph{climbing Jacob's ladder} for the effort of developing more advanced xc functionals to reach chemical accuracy.\cite{Pedrew2001JacobsLadder}
Higher rungs offer higher accuracy at the price of higher computational costs and lower transferability.
Novel xc potentials based on machine learning \cite{Nagai2020xcML,Burke2022DFTML} attempt to address the accuracy issue but still struggle with lower transferability.
It remains to be seen whether machine learning functionals will replace traditional ones in the coming years.
Machine-learned xc potentials as well as machine-learned force fields \cite{Behler-MLFF2007, BerkelbachAFQMCFF2022} require highly accurate reference data.
In addition, more accurate methods are also important to calibrate new methods or to access strongly correlated materials where DFT generally performs poorly.

For this reason, a whole spectrum of correlation-consistent methods has been developed over the last 50 years.\cite{SzaboQC1996,Helgaker2000QC} Full configuration interaction (FCI) provides an exact solution within a given basis set \cite{Handy1980FCI} but suffers from exponential scaling. Therefore, it is only usable for modest system sizes (up to $10^{10}$ Slater determinants). A clever partitioning of the Fock space led to a class of methods called selected CI \cite{Umrigar2016HeatBath,Sharma2017HeatBath,Dash2018CIPSI} and its stochastic counterpart full configuration interaction quantum Monte Carlo (FCIQMC). \cite{Booth2009FCIQMC,Cleland2010IFCIQMC,Ghanem2020ASFCIQMC} These methods enabled the treatment of the Fock space with up to $10^{30}$ Slater determinants while maintaining the accuracy of the traditional FCI.
However, they still scale weakly exponentially making them difficult to apply to more realistic systems. 

Besides CI methods, the most prominent methods are based on the coupled-cluster expansion. The most popular among them, the "gold-standard" coupled-cluster singles, doubles, and perturbative triples (CCSD(T)),\cite{Bartlett1982CCSD,Bartlett2007CC} is extensively used to treat molecules and is known to give excellent results for systems with small static correlation effects. High-accuracy benchmark data for small molecules are usually calculated using the CC expansion. For instance, a series of papers focused on high-accuracy extrapolated ab-initio thermochemistry (HEAT) include electron-electron correlation up to full quintuples in the coupled cluster expansion.\cite{Tajti2004Heat,Bomble2005Heat,Harding2008Heat} For the molecules in the HEAT set, these results are considered to be as good as the FCI results. More importantly, the HEAT studies also show that CCSD(T) alone is not accurate enough to achieve chemical accuracy ($< 1 \text{kcal}/ \text{mol}$) but approaches it in many relevant cases. Furthermore, CCSD(T) scales adversely with the 7th power of the system size and performs poorly in the presence of strong static correlation effects, e.g. bond dissociation. These drawbacks emphasize the need for alternative methods.

The density-matrix renormalization group (DMRG) is a variational method widely used to treat systems with strong correlation effects.\cite{White1999DMRG,Schollwock2005DMRG,Chan2011DMRGRev}
It uses matrix-product states to encode the locality in one of the spatial dimensions and to reduce the exponential scaling of the CI expansion. For low-dimensional quantum systems (quantum dots, molecules extended in one dimension), DMRG is the method of choice due to its efficiency and accuracy.
Although DMRG is applicable to arbitrary systems, its limitation of about 30 active orbitals prevents it from being a general-purpose tool for more generic and compact molecules or solids.

Quantum Monte Carlo methods, such as variational Monte Carlo \cite{McMillan1965VMC} (VMC) and diffusion Monte Carlo \cite{Anderson1976DMC,Ceperley1977DMC,Foulkes2001QMC} (DMC), exhibit low polynomial scaling with system size and are nowadays routinely applied to systems with hundreds of electrons. However, Nemec and co-workers\cite{Nemec2010DMC} showed that achieving chemical accuracy is difficult with DMC. They reported a mean absolute deviation of atomization energies of $3.2 \text{kcal}/\text{mol}$ for 55 molecules in the G2 set.\cite{Curtiss1991G2} Additionally, VMC and DMC require accurate models for multi-electron wavefunctions including linear combinations of Slater determinants, Jastrow factors, backflow wavefunctions, and many others.
Finally, recent work that combines Deep Learning and VMC has emerged as a promising way to tackle the quantum many-body problem.\cite{Pfau2020Ferminet,Spencer2020Ferminet,Scherbela2022DeepErwin}

In this work, we use the auxiliary-field quantum Monte Carlo (AFQMC) method. AFQMC is a well-established projector Monte Carlo method successfully used in various applications. \cite{Al-Saidi2007BondBreaking,Purwanto2009Excited,Purwanto2011HCa,Virgus2012CobaltGraphene,Purwanto2105ChromiumDimer,Motta2017Hchain,Lee2020BenzeneAFQMC,Lee2020SymmetryBreaking,Rudhsteyn2022Metallocene} The original AFQMC was formulated as a path-integral method using the Metropolis algorithm.\cite{Hirsch1985Hirsch-Fye,Sugiyama1986AFQMC,White1989AFQMC,Sorella1989AFQMC} It behaved well for systems without severe fermionic sign problem. Since almost all general fermionic systems suffer from the sign problem, the reformulation into an open-ended random walk led to the constrained path (cp)-AFQMC, which was remarkably successful for model Hamiltonians.\cite{Zhang1995Constrained,Zhang1997Constrained} In order to handle general many-body Hamiltonians, the phaseless (ph)-AFQMC was proposed by Zhang and co-workers.\cite{Zhang2003Phaseless} The ph-AFQMC was successfully applied to isolated molecules using Gaussian-type orbitals (GTOs) and to solids using a plane-wave basis (PWs).\cite{Suewattana2007Solids} With the increasing popularity of the method, a wide range of ph-AFQMC applications emerged in recent years.\cite{Al-Saidi2007BondBreaking,Purwanto2009Excited,Purwanto2011HCa,Virgus2012CobaltGraphene,Purwanto2105ChromiumDimer,Motta2017Hchain,Lee2020BenzeneAFQMC,Lee2020SymmetryBreaking,Rudhsteyn2022Metallocene} One reason for its popularity is that the ph-AFQMC scales quartic with the system size. This scaling stems from the local energy evaluation and can be further reduced using tensor contraction,\cite{Motta2109-AFQMC-ED,Lee2020AFQMCreduced} density-fitting techniques \cite{Malone2019AFQMC-DF,Weber2022LOAFQMC} or plane-wave basis.\cite{Suewattana2007Solids}
In contrast to DMC, AFQMC yields energies fully consistent with other traditional quantum chemistry approaches, allowing correlation energies to be directly compared.

Controlling the fermionic phase problem is crucial for the accuracy of the AFQMC method. 
In the ph-AFQMC, the \emph{phaseless approximation} imposes a constraint on the walker weights,\cite{Zhang2003Phaseless,Zhang2013Juelich,Motta2018AFQMCREVIEW} analogous to the fixed-node approximation in DMC.\cite{Anderson1976DMC}
The phaseless approximation can introduce somewhat uncontrolled errors that often lead to overestimated correlation energies.
In the ph-AFQMC, improving the trial wavefunctions systematically reduces the phaseless approximation error.
As the trial wavefunction approaches the exact ground-state wavefunction, the ph-AFQMC energy approaches the exact ground-state energy and the phaseless error disappears.
Usually, the trial wavefunction consists of the single Slater determinant formed from Hartree-Fock (HF) or Kohn-Sham (KS) orbitals.
They are particularly popular because they are easily accessible from all quantum chemistry / solid-state physics codes and avoid additional computational costs introduced by more complex trial wavefunctions.
Recent research focused on multi-determinant trial wavefunctions\cite{Shi2013MDWF,Chang2016TrialWF,Mahajan2021TamingSignProblem,Mahajan2022AFQMCsCI,Huggins2022qcAFQMC}.
For example, Mahajan \emph{et. al.} used $10^4$ Slater determinants while increasing the total cost of the ph-AFQMC computation by only a factor of 3 in comparison to the single Slater determinant case.

In this study, we adopt a simpler approach. We will carefully scrutinize whether modifications to the weight update reduce the need to go beyond single-determinant trial wavefunctions. With these modifications, the ph-AFQMC can provide close to chemical accuracy for the HEAT set. Bomble \emph{et. al.}\cite{Bomble2005Heat} provided highly accurate CCSDTQP molecular energies at the double-zeta basis set (cc-pVDZ). In this work, we benchmark the accuracy of the ph-AFQMC energies against this reference data set of small molecules.
Borda \emph{et. al.} recently performed a similar study on the G1 test set.\cite{Borda2019NONSD}

The rest of the paper is structured as follows: In Sec.~\ref{theory}, the ph-AFQMC method is briefly reviewed along with the proposed modifications, followed by details on the implementation in Sec.~\ref{compdet}. In Sec.~\ref{results}, we present and discuss different applications of ph-AFQMC. Finally, Sec.~\ref{conclusion} concludes our findings and possibilities for future developments.

\section{AFQMC Formalism} 
\label{theory}
In this section, we briefly introduce the AFQMC formalism.
For a more detailed overview of the theory, we refer the interested reader to one of the reviews.\cite{Zhang2003Phaseless,Zhang2013Juelich,Motta2018AFQMCREVIEW} 
Consider the full many-body Hamiltonian written in any orthonormal one-particle basis given by
\begin{align}
\begin{split}
    \hat H  & = \hat H_1 \ + \ \frac{1}{2} \sum_{g} \hat L_{g}^{2}   \\
            & = \sum_{pq} h_{pq} \; \hat a_{p}^{\vphantom{\dagger}\dagger} \hat a_{q} \ + \ \frac{1}{2} \sum_{g} \ \sum_{pq} L_{g,pq} \; \hat a_{p}^{\vphantom{\dagger}\dagger} \hat a_{q} \ \sum_{rs}  L_{g,rs} \; \hat a_{r}^{\vphantom{\dagger}\dagger} \hat a_{s},
\end{split}
\end{align}
where $\hat a_{p}^{\dagger}$ and $\hat a_{q}$ are  fermionic creation and annihilation operators, respectively. The single-particle Hamiltonian matrix elements $h_{pq}$ include all one-body terms. The two-body Hamiltonian is written as a sum of squares of one-body operators $\hat L_g$. In the Gaussian basis, we can obtain these operators by iterative Cholesky decomposition \cite{Beebe1977CholDec,Koch2003CholDec} of electron repulsion integrals (ERIs). The indices $p$, $q$, $r$, and $s$ go over $N$ basis functions and the index $g$ goes over $N_g$ Cholesky vectors. Typically, $N_g \approx 10 N$. We neglect the spin indices for simplicity. 

The exact many-body ground-state wavefunction $\ket{\Phi_0}$ is extracted from the long-time imaginary propagation of an initial state $\ket{\Psi_{\init}}$
\begin{equation}
    \ket{\Phi_0} = \lim_{n \to \infty} \left[ e^{-\tau \left( \hat H - E_0 \right) }  \right]^{n} \ket{\Psi_{\init}},
\end{equation}
where $\tau$ is the imaginary time step and $E_0$ is the estimated ground-state energy.
This equation is exact in the limit of infinitesimally small time steps. 
In this work, the Hartree-Fock orbitals form the initial wavefunction. 

To treat the electron-electron interaction efficiently, the Hubbard-Stratanovich (HS) transformation \cite{Hubbard1959,Stratonovich1957} is used
\begin{equation}
\label{eq:hstrafo}
    e^{-\frac{\tau}{2} \sum_g \hat L_{g}^{2}} = \int \text{d}x^{N_g} \; p(x_g) \; e^{i \sqrt{\tau} \sum_g x_g \hat L_g} + \mathcal{O}(\tau^2)~, 
\end{equation}
where $p(x_g)$ is the standard normal distribution and $x_g$ are random numbers drawn from this distribution. 
In other words, the Hubbard-Stratanovich transformation maps the actual system of interacting particles onto a system of non-interacting particles coupled to random fields. Since ERIs are positive definite, random fields are purely imaginary, i.e., one can define a complex-valued effective Hamiltonian $\bar h$ given by
\begin{equation}
\label{eq:heff}
    \bar h_{pq}^{w} = h_{pq} - \frac{i}{\sqrt{\tau}} \sum_{g} x_g^{w} L_{g,pq},
\end{equation}
where the superscript $w$ emphasizes that unique random fields are drawn for each walker.
The ensemble of $N_w$ walkers represents the exact many-body ground state, where each walker is represented by the following elements: real-valued weight $W$, phase $\theta$, and single Slater determinant $\ket{\Psi}$. 
Therefore, the exact many-body ground-state wavefunction can be approximated as 
\begin{equation}
\label{eq:wavefun_rep}
    \ket{\Phi_0} = \Bigl(\sum_{kw} W_{k}^{w} e^{i \theta_{k}^{w}}\Bigr)^{-1} \sum_{kw} W_{k}^{w} e^{i \theta_{k}^{w}} \frac{\ket{\Psi_{k}^{w}}}{\braket{\Psi_{\trial} | \Psi_{k}^{w}}},
\end{equation}
where the index $k$ stands for time steps and the index $w$ for walkers.
$\ket{\Psi_{\trial}}$ denotes the trial wavefunction. 
The walkers are initialized as follows:
\begin{align}
    W_{0}^w &= 1; 
    &
    \theta_{0}^w &= 0;
    &
    \ket{\Psi_{0}^w} &= \ket{\Psi_{\init}}.
\end{align}
The equations of motion for the walkers are
\begin{gather}
    \ket{\Psi_{k+1}^w} = e^{- \tau \bar h^{w}} \ket{\Psi_{k}^w},    \label{eq:psiupdate}    \\
    W_{k+1}^w e^{i \theta_{k+1}^w} = W_{k}^w e^{i \theta_{k}^w} \; \frac{\braket{\Psi_{\trial} | \Psi_{k+1}^w}}{\braket{\Psi_{\trial} | \Psi_{k}^w}}.   \label{eq:wupdate}
\end{gather}
During the AFQMC propagation, this approximate ground-state wavefunction grants access to the observables of the system. The simplest and probably the most important observable---the total energy---is defined as the mixed expectation value of the Hamiltonian
\begin{equation}
\label{eq:ensamp}
    E_0 = \frac{\braket{\Psi_{\trial}|\hat H|\Phi_0}}{\braket{\Psi_{\trial}|\Phi_0}} \approx \frac{\sum_{kw} W_{k}^w e^{i \theta_{k}^w} E_{\text{loc}}(\Psi_{k}^w)}{\sum_{kw} W_{k}^w e^{i \theta_{k}^w}},
\end{equation}
where the local energy estimator $E_{\text{loc}}(\Psi^{w})$ is computed using the generalized Wick's theorem \cite{Balian1969Wick}
\begin{multline}
\label{eq:locen}
    E_{\text{loc}}(\Psi^{w}) = \frac{\braket{\Psi_{\trial} | \hat H | \Psi^{w}}}{\braket{\Psi_{\trial} | \Psi^{w}}}  \\
    = \sum_{pq} h_{pq} G_{pq}^{w} + \frac{1}{2} \sum_g \sum_{pqrs} L_{g,pq} L_{g,rs} (G_{pq}^{w}G_{rs}^{w} - G_{ps}^{w}G_{rq}^{w}).
\end{multline}
The one-body reduced density matrix $G_{pq}^{w}$ is defined as 
\begin{equation}
    G_{pq}^{w} \equiv G_{pq}(\Psi^{w}) = \frac{\braket{\Psi_{\trial} | \hat a_{p}^{\vphantom{\dagger}\dagger} \hat a_q| \Psi^{w}}}{\braket{\Psi_{\trial} | \Psi^{w}}} = \left[ \Psi^{w} (\Psi_{\trial}^{\dagger} \Psi^{w})^{-1}  \Psi_{\trial}^{\dagger} \right]_{qp}.
\end{equation}
Here, $\Psi_{\trial}$ and $\Psi^{w}$ represent the $N_{\mathrm e}$ occupied orbitals in the $N$ basis functions. Equations (\ref{eq:psiupdate}, \ref{eq:wupdate}, \ref{eq:ensamp}, \ref{eq:locen}) define the free-propagation (fp)-AFQMC.

\subsection{Mean-Field Subtraction}
A decrease in the magnitude of the diagonal components of the Cholesky vectors $L_{g,pq}$ leads to smaller fluctuations in AFQMC and thus better statistics. To this end, one introduces a shift
\begin{equation}
    \bar L_g = \braket{\hat L_g}_{\trial} = \frac{\braket{\Psi_{\trial}|\hat L_g | \Psi_{\trial}}}{\braket{\Psi_{\trial}| \Psi_{\trial}}}   
\end{equation}
to the two-body Hamiltonian
\begin{equation}
    \hat H_2 = \frac{1}{2} \sum_g \left( \hat L_g - \bar L_g \right)^2 + \sum_g \bar L_g \hat L_g - \frac{1}{2} \sum_g \bar L_g^2~.
\end{equation}
Here, we identify the second term as the classical Hartree potential corresponding to the trial wavefunction and add it to the one-body Hamiltonian.
The last term is the Hartree energy corresponding to the trial wavefunction. 
The mean-field subtraction incurs no computational cost for the AFQMC propagation.

\subsection{Importance Sampling}
Introducing an importance function in AFQMC shifts the Gaussian probability density in Eq.~\eqref{eq:hstrafo}. This shift is referred to as the "force bias". Choosing the force bias as 
\begin{equation}
\label{eq:forcebias}
    f_{g}^{w} = - i \sqrt{\tau} \ \frac{\braket{\Psi_{\trial} | \hat L_g - \bar L_g | \Psi^{w} }}{\braket{\Psi_{\trial} | \Psi^{w}}}.
\end{equation}
minimizes the statistical fluctuations in the random fields.\cite{Zhang2013Juelich} After applying the mean-field subtraction and importance sampling, the effective Hamiltonian (Eq.~\eqref{eq:heff}) changes to
\begin{equation}
\label{eq:heffTOT}
    \bar h_{pq}^{w} = h_{pq} - \frac{i}{\sqrt{\tau}} \sum_{g} (x_{g}^{w} - f_{g}^{w})  (L_{g,pq} - \bar L_g \delta_{pq}).
\end{equation}
The force bias is a walker-dependent quantity and must be calculated in each time step for each walker. The force bias is crucial to successful AFQMC simulations, however, computing it entails considerable computational effort for the AFQMC procedure and may even be the computational bottleneck.\cite{Motta2018AFQMCREVIEW}

\subsection{Phaseless Approximation}
\label{phaseless}

The fp-AFQMC is formally exact but suffers from the fermionic phase problem.  The problem is observed in the expression for the energy evaluation (Eq.~\eqref{eq:ensamp}). After the equilibration time, during which excited states contained in $\ket{\Psi_{\init}}$ decay exponentially, walkers will populate the entire complex plane uniformly. Therefore, the denominator in Eq.~\eqref{eq:ensamp} vanishes and the whole expression becomes ill-defined. To circumvent the fermionic phase problem, the phaseless approximation is introduced.

Since the effective Hamiltonian given in Eq.~\eqref{eq:heffTOT} is complex-valued, the orbitals become complex-valued as well. In DMC, $\ket{\Phi}$ and $-\ket{\Phi}$ are both valid solutions of the Schr\"odinger equation. Similarly in AFQMC, if $\ket{\Phi}$ is the valid solution, then $e^{i \theta} \ket{\Phi}$ is a valid solution for any $\theta \in [0, 2\pi)$. To prevent the exponential growth of the noise in AFQMC simulation, Zhang \emph{et. al.}\cite{Zhang2003Phaseless} introduced a phaseless approximation
\begin{gather}
    \begin{aligned}
    \label{eq:phaseless}
    W_{k+1}^w = W_{k}^w \: \left | \frac{\braket{\Psi_{\trial} | \Psi_{k+1}^w}}{\braket{\Psi_{\trial} | \Psi_{k}^w}} I^{w} \right| \ \text{max}\left( 0, \: \text{cos}(\Delta \theta) \right),
    \end{aligned}   \\
    \theta_{k}^w = 0,
\end{gather}
where the importance sampling factor $I^{w}$ and the phase change $\Delta \theta$ are defined as  
\begin{gather}
    I^{w} = \exp\Bigl[\sum_g x_g^w f_g^w - \frac{1}{2} f_g^w f_g^w\Bigr], \\
    \Delta \theta = \mathfrak{Im} \ \text{ln} \frac{\braket{\Psi_{\trial} | \Psi_{k+1}^w}}{\braket{\Psi_{\trial} | \Psi_{k}^w}} \approx \mathfrak{Im} \sum_{g} x_g^w f_g^w .
\end{gather}
The update equation~\eqref{eq:phaseless} can be also rewritten as 
\begin{equation}
    W_{k+1}^w = W_{k}^w \: \text{max} \left(0, \ \mathfrak{Re} \frac{\braket{\Psi_{\trial} | \Psi_{k+1}^w}}{\braket{\Psi_{\trial} | \Psi_{k}^w}} \right) \: |I^w|.
\end{equation}
Zhang \emph{et. al.} \cite{Zhang2003Phaseless, Zhang2013Juelich} state that different approximations yield the same expectation values and 
slightly different standard deviations.

In this work, we investigate this statement more quantitatively. To this end, we modify the expression for the total energy evaluation (Eq.~\eqref{eq:ensamp})
\begin{equation}
    E_0 = \frac{\sum_{kw} W_{k}^w \cos (\theta_{k}^w) E_{\text{loc}}(\Psi_{k}^w)}{\sum_{kw} W_{k}^w \cos (\theta_{k}^w) }
\end{equation}
and the update procedure for the walker weights
\begin{equation}
\begin{aligned}
\label{eq:phaseless*}
    W_{k+1}^w e^{i \theta_{k+1}^w} = W_{k}^w e^{i \theta_{k}^w} \: \frac{\braket{\Psi_{\trial} | \Psi_{k+1}^w}}{\braket{\Psi_{\trial} | \Psi_{k}^w}} \mathfrak{Re} I^w~.
\end{aligned}
\end{equation}
We refer to this new approach as ph$^\ast$-AFQMC throughout this paper.
In contrast to the original approach, there are two key differences: 
First, we retain the complex nature of the walker weights and use the real part for the population control and evaluation of physical observables, such as the energy. Walkers are killed explicitly if $|\theta_{k}^w| \geq \frac{\pi}{2}$.
Secondly, we include the real part of the importance weight instead of the absolute value. We suggest that only the real part is meaningful and using the absolute value increases the weights with the imaginary part. This may contribute to the systematic over-correlation of ph-AFQMC. We will show in Sec.~\ref{results} that ph$^\ast$-AFQMC systematically yields correlation energies smaller in absolute value.

\section{Implementation Details}
\label{compdet}

In this section, we present our implementation of AFQMC in QMCFort.\footnote{QMCFort is available from the authors upon a reasonable request.}
QMCFort is written in Fortran and utilizes (threaded) BLAS/LAPACK for fast linear algebra and MPI and OpenMP for parallelization.
In addition to AFQMC, \mbox{QMCFort} offers implementations of restricted, 
restricted open-shell, and unrestricted Hartree-Fock (RHF, ROHF and UHF) and MP2. 
The electron repulsion integrals (ERIs) are calculated using the McMuchie-Davidson scheme\cite{Helgaker2000QC} and decomposed into Cholesky vectors.
Therefore, QMCFort can act as a standalone tool to setup and run AFQMC simulations.
However, only the AFQMC part is highly optimized for large-scale calculations.
For this reason, we also interface QMCFort with PySCF\cite{pyscf} for isolated systems and with VASP\cite{Kresse1996Vasp} for extended systems. 
The implementation of the computationally intensive parts of the AFQMC procedure and the interface between QMCFort and VASP are detailed in the remainder of this section.

\subsection{Effective Hamiltonian}
\label{effham}
At each time step and for each walker, we build the effective Hamiltonian according to Eq.~\eqref{eq:heffTOT}.
The computationally demanding part is the convolution of the shifted random fields with Cholesky vectors $\sum_g (x_g^w - f_g^w) \cdot L_{g,pq}$.
Treating the orbital indices $p$ and $q$ as a combined index, we map this contraction to matrix-matrix multiplication with a cubic scaling ($\sim N^2N_gN_w$) in system size.
Typically, we treat in the range of 10 to 100 walkers per MPI rank and compute this contraction for all of them simultaneously.
Although the exchange energy evaluation (Sec.\ref{exchange}) scales worse, the creation of the effective Hamiltonian is often the most expensive operation in AFQMC because it is required at each time step. 

\subsection{AFQMC Propagation}
\label{afqmcprop}
The effective Hamiltonian propagates the orbitals according to Eq.~\eqref{eq:psiupdate}.
The matrix exponentials in the Trotter-Suzuki propagator \cite{Trotter1959,Suzuki1976} are approximated by the sixth-order Taylor expansion. 
This requires several applications of the effective Hamiltonian to the orbitals, i.e. matrix-matrix multiplications of the form
\begin{equation}
    \Psi^{w}_{pi} \to \bar \Psi^{w}_{pi} = \sum_{q} \bar h_{pq}^{w} \Psi^{w}_{qi}~.
\end{equation}
These scale as $N^2N_e$.

\subsection{Force Bias}
\label{forcebias}
The force bias evaluation (see Eq.~\eqref{eq:forcebias}) is equivalent to
the evaluation of the Hartree potential and energy.
It scales cubically with respect to the system size, i.e. $NN_gN_eN_w$.
Similar to the effective Hamiltonian (Sec.~\ref{effham}), the force bias is computed for all walkers simultaneously.
To make it more transparent, we will first define the contracted Cholesky vectors
\begin{equation}
    \label{eq:contcholesky}
    \mathcal{L}_{g,pi} = \sum_{q} L_{g,pq} \Psi_{\text{T},qi},
\end{equation}
and the force bias is then
\begin{equation}
    f_{g}^{w} = \sum_{pi}  \mathcal{L}_{g,pi} \Psi_{pi}^{w}~.
\end{equation}
Here, we can treat $\{p, i\}$ as a single index and evaluate the contraction with a matrix-matrix multiplication.
We use the force bias to compute the Hartree energy for the walker $\Psi^w$
\begin{equation}
    E_{\text{H}}(\Psi^w) = \frac{1}{2} \sum_g f_g^w f_g^w.
\end{equation}

\subsection{Exchange Energy}
\label{exchange}
Of all individual contributions to AFQMC, the computation of the exchange energy incurs the largest computational cost.
Although the naive implementation of Eq.~\eqref{eq:locen} scales as $N^3N_gN_w$, we simplify it by computing intermediate arrays $\alpha_{g,ij}^{w}$
\begin{equation}
    \alpha_{g,ij}^{w} = \sum_p \mathcal{L}_{g,pi} \Psi_{pj}^{w}~.
 \end{equation}
The exchange energy of the walker $\Psi^w$ is then given by
\begin{equation}
    E_{\text{x}}(\Psi^w) = -\frac{1}{2} \sum_g \sum_{ij} \alpha_{g,ij}^{w} \alpha_{g,ji}^{w}~.
\end{equation}
The quartic-scaling ($NN_gN_e^2N_w$) construction of the intermediate arrays $\alpha_{g,ij}^{w}$ is also mapped to a matrix-matrix multiplication.
Multiple walkers are treated simultaneously in a block of approximately 5-10 walkers for optimal performance.
One can evaluate the exchange energy only say every 10th time step.
This has no significant impact on the statistics because one needs to sample independent local energies so the alternative is block averaging.
Computing the exchange energy from the $\alpha$ tensors scales
as $N_gN_e^2$ but the computational effort is negligible in comparison to other tasks discussed in this section. 
\subsection{Other tasks}

Besides the operations mentioned above, there are a few additional parts of the AFQMC procedure that should be mentioned.
Since the AFQMC propagator is non-unitary, we periodically reorthogonalize the walkers by a QR decomposition of the walker matrices $\Psi^w$. 
The walker population is controlled to avoid too small or too large weights. 
We opted for the comb method\cite{Calandra1998POP}, because it keeps the total number of walkers constant. 

\subsection{Performance}
\label{performance}

\begin{table}
\caption{\label{tab:perf} Number of operations per walker and time step and asymptotic scaling of the computationally intensive parts of the AFQMC procedure.
$N$ represents the number of basis functions, $N_g$ is the number of Cholesky vectors, and $N_e$ is the number of occupied states in the system.}
\begin{ruledtabular}
\begin{tabular}{lcc}
Task                   & \multicolumn{1}{c}{\# of operations} & \multicolumn{1}{c}{asymptotic scaling} \\ \hline
effective Hamiltonian  & $N^2N_g$    & $\mathcal{O}(N^3)$   \\
AFQMC propagation      & $N^2N_e$     & $\mathcal{O}(N^3)$  \\
force bias             & $NN_gN_e$    & $\mathcal{O}(N^3)$  \\
exchange energy        & $NN_gN_e^2 + N_gN_e^2$   & $\mathcal{O}(N^4)$   \\ 
\end{tabular}
\end{ruledtabular}
\end{table}

In Table \ref{tab:perf}, we summarize the number of operations corresponding to the computationally demanding tasks.
Assuming that the number of walkers is independent of the system size, we list the resulting asymptotic large system-size scaling for these tasks.
We gauge the asymptotic behavior for water clusters of various sizes containing up to 5 water molecules.
The calculations were performed on a dual-socket Intel Skylake Platinum 8174 with 24 cores each and a nominal base frequency of \SI{3.1}{\giga \hertz}. 
We average the timings over 200 AFQMC steps, with each step processing 4\,800 walkers on 48 MPI processes.

\begin{figure}
    \centering
    \includegraphics[width=\columnwidth]{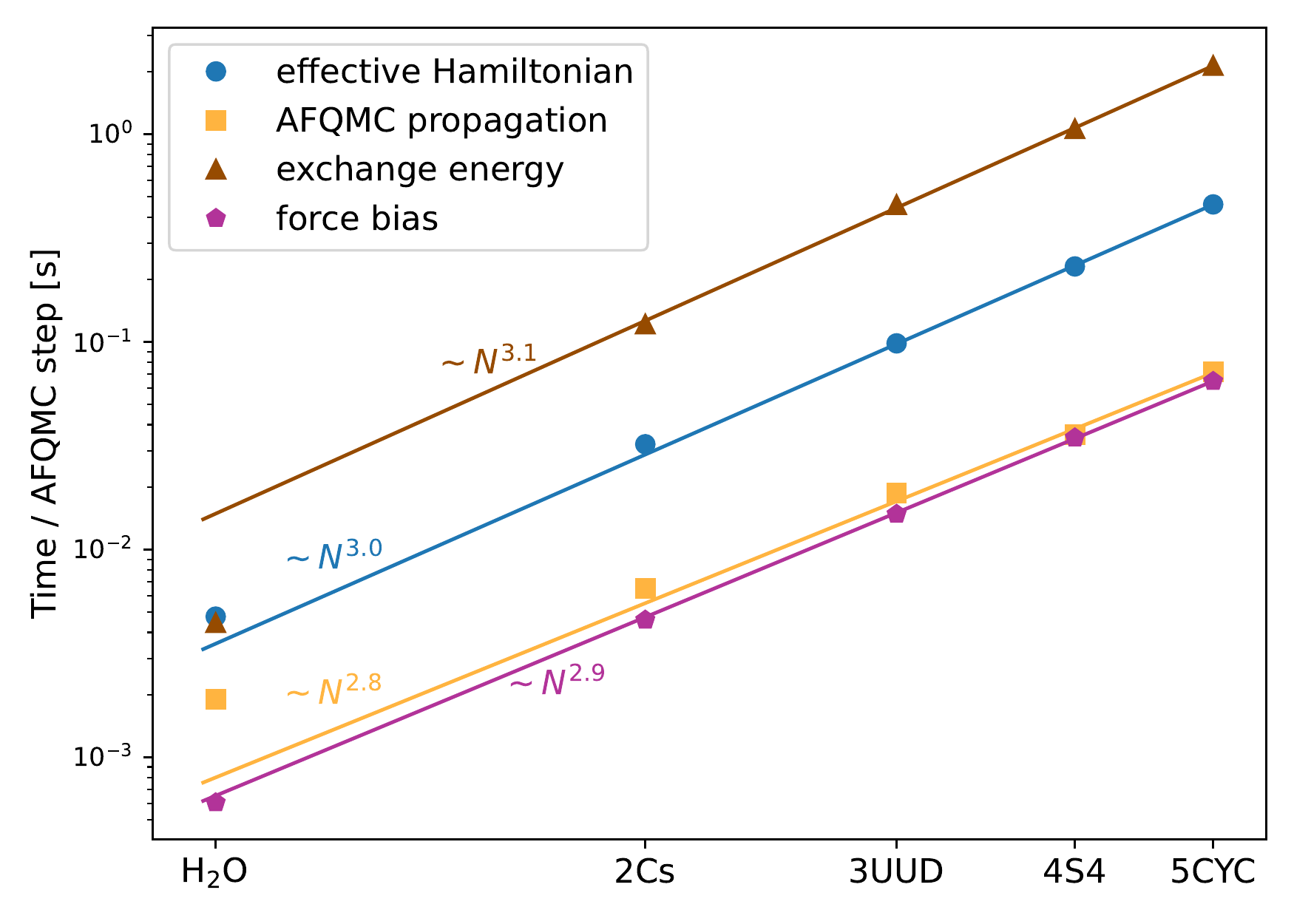}
    \caption{Elapsed time per AFQMC step of the computationally intensive operations measured on water clusters of different sizes.
    AFQMC calculations were performed on a dual-socket Intel Skylake Platinum 8174  with 4\,800 walkers, and 48 MPI processes.
    Timings are averaged over 200 AFQMC steps.
    }
    \label{fig:tim}
\end{figure}

\begin{figure}
    \centering
    \includegraphics[width=\columnwidth]{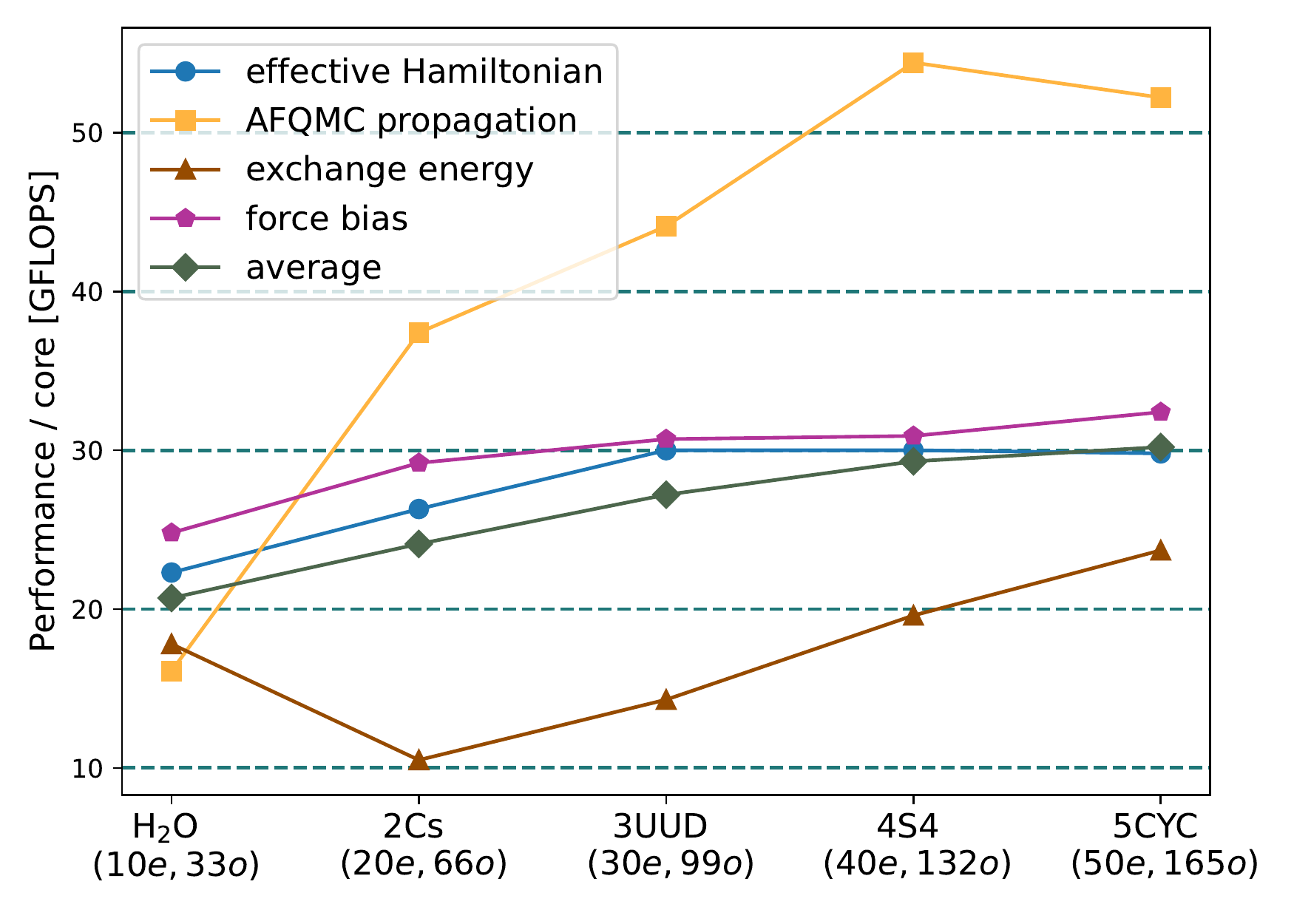}
    \caption{Performance per CPU core of the computationally intensive operations in the AFQMC procedure measured on water clusters of different sizes. 
    Tuples on the $x$-axis denote the number of electrons and orbitals. 
    Executing the AFQMC code on one CPU node (dual-socket Intel Skylake Platinum 8174) delivers 1--1.5~TFLOPS (30~GFLOPS per core, green line) on average.}
    \label{fig:perf}
\end{figure}

Fig.~\ref{fig:tim} shows the elapsed times per AFQMC step of the computationally demanding tasks.
To estimate the effective scaling exponent for each task, we fit the power function $f(x) = C x^{\beta}$ through the data points.
As predicted, the construction of the effective Hamiltonian, AFQMC propagation, and force bias calculation scale cubically.
The measured exponent $\beta=3.1$ for the exchange energy does not exactly match the expected exponent $\beta = 4$.
With increasing system size, we observe an increasing per-core performance of the exchange evaluation (Fig.~\ref{fig:perf}).
This increase effectively reduces the scaling exponent.
Still, the exchange evaluation is the least performant of the relevant computational tasks.
As depicted in Fig.~\ref{fig:perf}, most other tasks run efficiently with the performance of 20--35 GFLOPS/core. 
A notable exception is the AFQMC propagation that reaches a performance above 50 GFLOPS/core for large systems, very close to peak performance.

\subsection{Interface with VASP}
\label{diamond}

VASP provides a set of one-electron mean-field orbitals $\{ \phi_p(\textbf{r}) \}$.
From these, we compute the single-particle Hamiltonian matrix elements $h_{pq}$ and Cholesky vectors $L_{g,pq}$ (see Appendix for more details).
We export the matrix elements and Cholesky vectors to a file and read them with QMCFort.
Then, the AFQMC calculation for extended systems is equivalent to calculations on isolated molecules using the Gaussian basis set.

\begin{figure}
    \centering
    \includegraphics[width=\columnwidth]{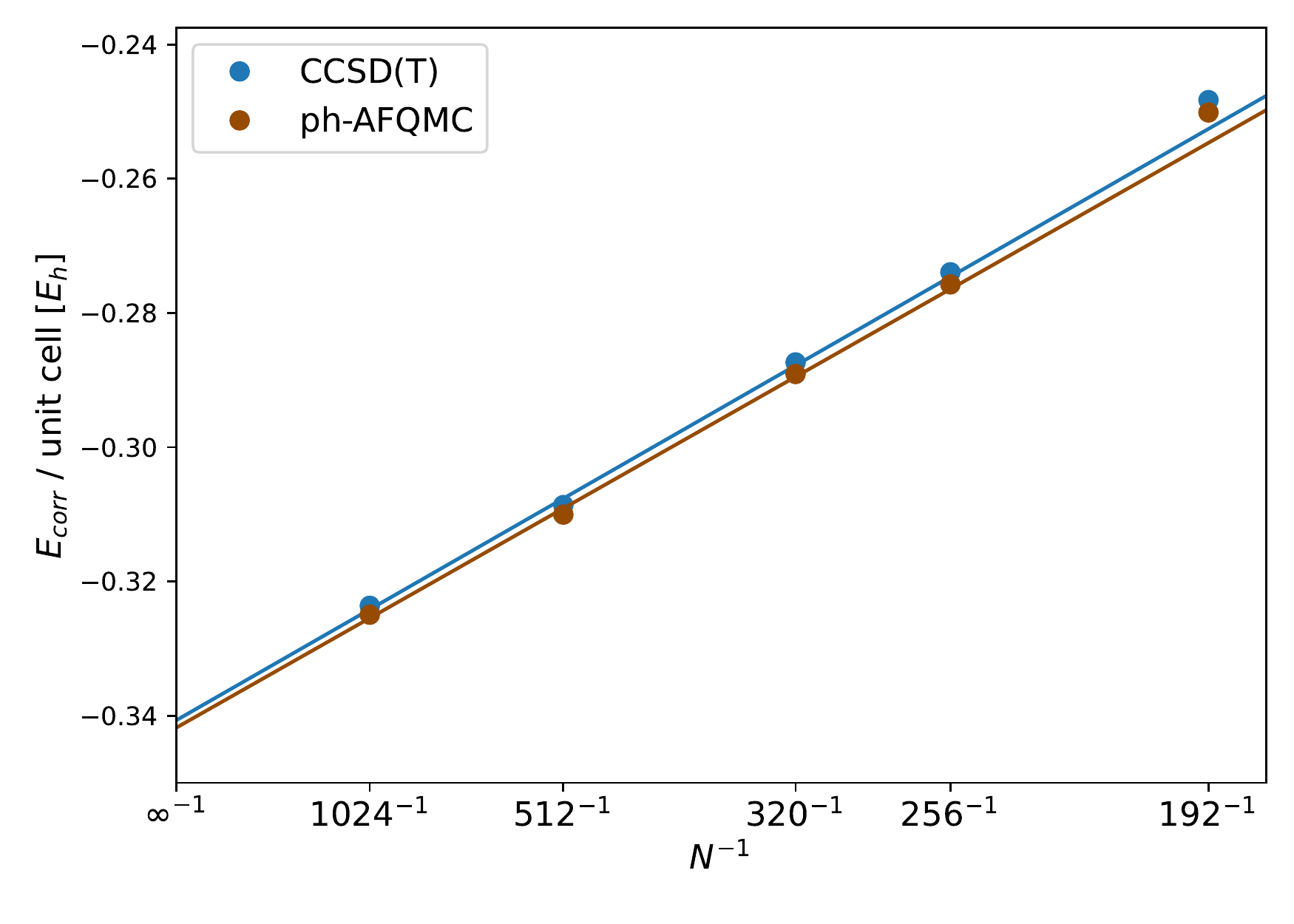}
    \caption{
    Convergence behavior of the CCSD(T) and ph-AFQMC correlation energies as a function of the number of canonical orbitals for diamond.
    Linear extrapolation to the complete basis set is performed using the first three points ($1024^{-1}$, $512^{-1}$, and $320^{-1}$).
    The calculations were performed using a 2x2x2 diamond supercell with an energy cutoff of $700\text{eV}$.
    }
    \label{fig:rpano}
\end{figure}

As an example, we compute the 2x2x2 supercell of diamond. 
The convergence of the AFQMC and CCSD(T) correlation energies with respect to the number of canonical orbitals is studied. 
CCSD(T) energies are computed using Cc4s code.\cite{HummelCc4s2017} 
A lattice constant of $a=3.567$\ \AA{} is used with an energy cutoff of $E_{\text{cut}} = 700\ \text{eV}$. 
In addition, we use an energy cutoff $E_{\text{cut}}^{\chi} = 500\ \text{eV}$ for the truncation of the Coulomb kernel.
VASP relies on PAW potentials\cite{blochl1994,Kresse1999PAW} and the potential referred to as C\_GW with valence $2s2p$ was used.
The core radius of this potential is $r_{\text{core}}=1.5\ \text{a.u.}$; the local potential is equivalent to the d-pseudopotential and 4 projectors are used (two $s$-projectors with $r_{\text{cut}}=1.2\ \text{a.u.}$, and two  $p$-projectors with $r_{\text{cut}}=1.5\ \text{a.u.}$).

Figure \ref{fig:rpano} shows the convergence of the correlation energy as a function of the number of canonical orbitals for ph-AFQMC and CCSD(T).
The number of orbitals is varied from the small basis set with 64 electrons in 192 orbitals (64e, 192o) to a relatively large basis of 64 electrons in 1024 orbitals (64e, 1024o).
A $1/N$ behavior is fitted through the data points.
The ph-AFQMC and CCSD(T) lines are almost indistinguishable suggesting that the methods yield consistent energies for periodic systems.
Further details on AFQMC calculations for extended systems will be reported in a forthcoming publication. \cite{AmirAFQMC}

To compare the computational cost of the CCSD(T) and ph-AFQMC, we conducted calculations using 256 and 1024 bands. For the 256-band calculation, the CCSD(T) calculation took 20 minutes on a single node (2x AMD Epyc (Milan)), while the ph-AFQMC calculation took 11 hours. However, for the 1024-band calculation, CCSD(T) required 24 hours on 4 nodes (2x AMD Epyc (Milan)), while ph-AFQMC calculation took 48 hours to complete. The timings we obtained for the CCSD(T) and ph-AFQMC calculations demonstrate the favorable scaling of the AFQMC method, which becomes increasingly advantageous as the number of orbitals and the complexity of the system increase.

\section{Results and Discussion}
\label{results}

\begin{table*}
\caption{\label{tab:heat_values} Total molecular energies (in Hartree units) for the HEAT set\cite{Tajti2004Heat} molecules at different levels of theory: ph-AFQMC\cite{Zhang2003Phaseless}, ph$^{*}$-AFQMC, CCSD(T), CCSDTQP\cite{Bomble2005Heat} and HCI\cite{Umrigar2016HeatBath,Sharma2017HeatBath}. All energies are calculated using the cc-pVDZ basis set. The last three rows of the table depict the root-mean-square deviation (RMSD), mean signed deviation (MSD), and mean absolute deviation (MAD).}
\begin{ruledtabular}
\begin{tabular}{l*{5}{d}}
Molecule            & \multicolumn{1}{c}{ph-AFQMC} & \multicolumn{1}{c}{ph$^{*}$-AFQMC} & \multicolumn{1}{c}{CCSD(T)} & \multicolumn{1}{c}{CCSDTQP} & \multicolumn{1}{c}{HCI} \\ \hline
H$_2$               & -1.16363(2)   & -1.16338(3)  & -1.163426   & -1.163426   & -1.163426   \\
CH                  & -38.37764(6)  & -38.37705(7) & -38.379655  & -38.380241  &             \\
CH$_2$              & -39.04181(5)  & -39.04150(6) & -39.041196  & -39.04165   &             \\
NH                  & -55.09087(5)  & -55.09077(7) & -55.097448  & -55.0917    &             \\
CH$_3$              & -39.71651(6)  & -39.71592(7) & -39.715543  & -39.71607   &             \\
NH$_2$              & -55.73265(6)  & -55.73222(8) & -55.732506  & -55.733076  &             \\
OH                  & -75.55854(6)  & -75.55837(8) & -75.559233  & -75.559689  &             \\
HF                  & -100.22933(7) & -100.2290(1) & -100.228131 & -100.228622 &             \\
H$_2$O              & -76.24249(8)  & -76.24167(9) & -76.241018  & -76.241649  & -76.241637  \\
NH$_3$              & -56.40334(7)  & -56.4024(1)  & -56.401913  & -56.402517  &             \\
C$_2$H              & -76.3999(2)   & -76.3986(2)  & -76.398558  & -76.401217  &             \\
CN                  & -92.4997(2)   & -92.4973(2)  & -92.488695  & -92.49276   & -92.492788  \\
C$_2$H$_2$          & -77.1118(2)   & -77.1092(2)  & -77.109249  & -77.110678  &             \\
CO                  & -113.0594(2)  & -113.0567(2) & -113.054431 & -113.055892 &             \\
HCN                 & -93.1912(2)   & -93.1881(2)  & -93.188321  & -93.18991   &             \\
N$_2$               & -109.2781(2)  & -109.2750(2) & -109.275298 & -109.277012 & -109.277005 \\  
HCO                 & -113.5778(2)  & -113.5757(2) & -113.575706 & -113.577384 &             \\
CF                  & -137.4764(1)  & -137.4754(2) & -137.474848 & -137.476019 &             \\
NO                  & -129.5970(2)  & -129.5945(2) & -129.597778 & -129.599737 &             \\
HNO                 & -130.1736(2)  & -130.1697(2) & -130.170989 & -130.172906 &             \\
O$_2$               & -149.9792(2)  & -149.9787(2) & -149.985684 & -149.987773 &             \\
HO$_2$              & -150.5606(2)  & -150.5603(2) & -150.558481 & -150.56038  &             \\ 
OF                  & -174.5004(2)  & -174.4998(2) & -174.497924 & -174.50009  &             \\
H$_2$O$_2$          & -151.1963(2)  & -151.1940(2) & -151.19363  & -151.195266 &             \\
F$_2$               & -199.0968(2)  & -199.0965(2) & -199.097448 & -199.099328 &             \\
CO$_2$              & -188.1561(2)  & -188.1525(2) & -188.147429 &             & -188.149551 \\ \hline 
RMSD (in \si{\milli \hartree}) &  2.89   &  2.73    & 1.65   & \\
MSD (in \si{\milli \hartree})  &  -0.38  &  1.17    & 1.39   & \\
MAD (in \si{\milli \hartree})  &  1.85   &  1.84    & 1.39    & \\
\end{tabular}
\end{ruledtabular}
\end{table*}

\begin{figure*}
    \centering
    \includegraphics[width=0.9\textwidth]{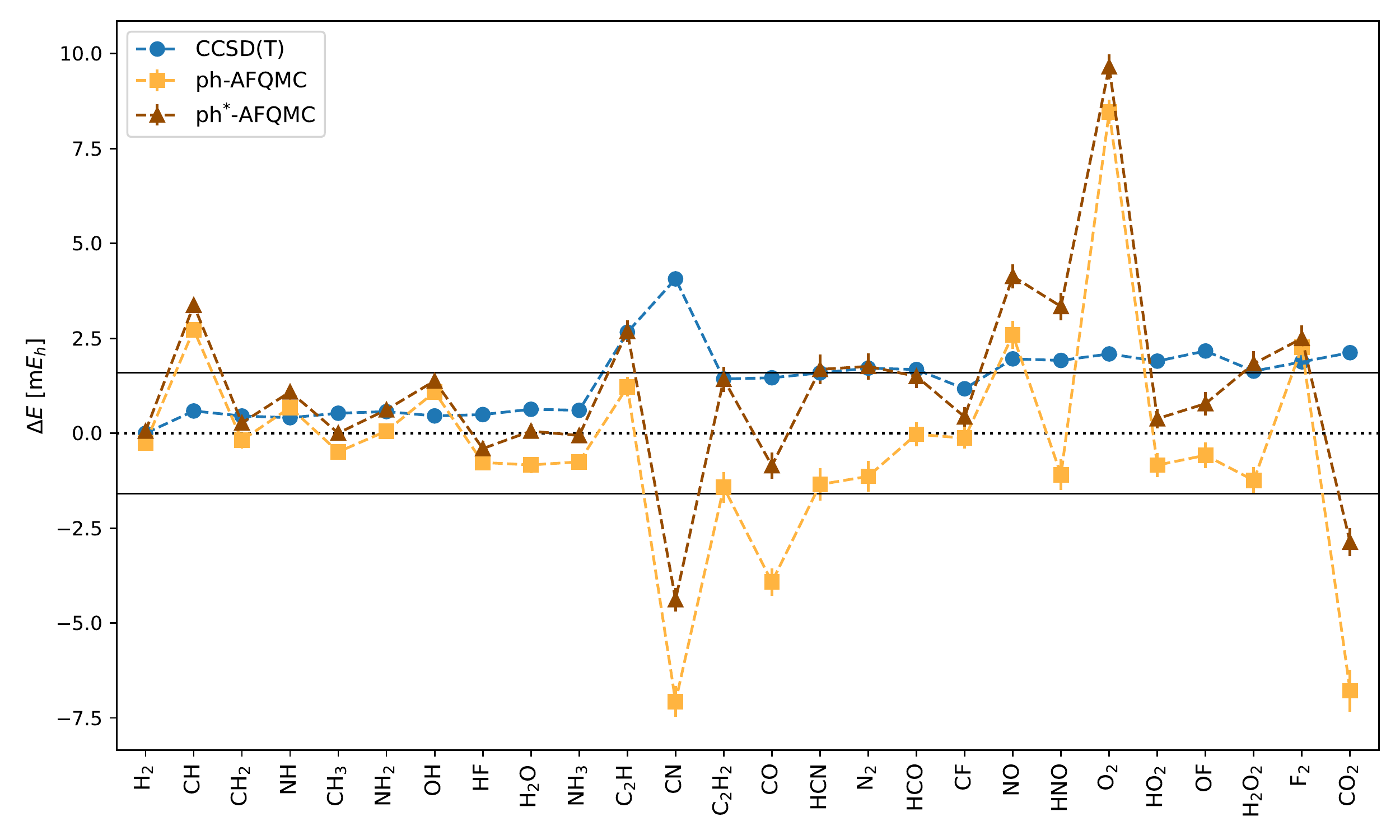}
    \caption{CCSD(T), ph-AFQMC and ph$^{*}$-AFQMC total-energy differences $\Delta E$ relative to the reference CCSDTQP energies for the HEAT set. Calculations use the cc-pVDZ basis set and a frozen-core approximation including only the last occupied shell. Trial wavefunctions are single RHF/UHF Slater determinants. The solid black line represents the chemical accuracy range ($\Delta E < \SI{1}{\kilo \cal / \mol}$)}
    \label{fig:heat}
\end{figure*}

In the following four subsections, we report (i) the total molecular energies for 26 molecules from the HEAT set, (ii) the AFQMC energy of the benzene molecule using the cc-pVDZ basis set, and (iii) the AFQMC binding energies of the water clusters. 

\subsection{Heat Molecules}
\label{HeatResults}
We calculated total energies for 26 molecules from the HEAT set using the frozen-core approximation and geometries from Ref.~\onlinecite{Bomble2005Heat} (HEAT).
All calculations employed the double-zeta basis set (cc-pVDZ).
We used the QMCFort code to calculate Cholesky vectors, the reference Hartree-Fock (RHF/UHF) wavefunctions, and all AFQMC energies. We used the PySCF package \cite{pyscf} to calculate the CCSD(T) energies.
The Dice code \cite{Umrigar2016HeatBath,Sharma2017HeatBath} was used to calculate the heat-bath CI energies.
To eliminate possible sources of systematic errors in the AFQMC, we (i) truncated Cholesky vectors at a conservative threshold of $10^{-8}$ to ensure a reasonably accurate representation of the molecular ERIs, (ii) chose a relatively small time step of \SI{0.002}{\hartree^{-1}} to exclude significant time-step errors, and (iii) equilibrated the system for 40\,000 time steps.
To ensure good statistics, we propagated 6\,000 walkers until the standard error of the mean of the predicted molecular energies dropped below \SI{0.2}{\milli\hartree}.

Table \ref{tab:heat_values} compares the ph-AFQMC and ph$^{*}$-AFQMC molecular energies to the "gold-standard" CCSD(T) and the more accurate coupled-cluster expansion including variational triple, quadruple and pentuple excitations (CCSDTQP).
We use the latter as reference values since they are practically converged against the FCI limit. 
The CCSDTQP value for the CO$_2$ molecule is missing in Ref.~\onlinecite{Bomble2005Heat}, so we use the result of the heat-bath configuration interaction (HCI).\cite{Umrigar2016HeatBath} 
To verify the accuracy of the HCI method, we selected a few other molecules (H$_2$, H$_2$O, CN, N$_2$) and found a maximal difference to the CCSDTQP values of \SI{0.01}{\milli \hartree}, well below the statistical AFQMC errors. 
Comparing to these reference values, we find a similar precision for the ph-AFQMC and ph$^{*}$-AFQMC method with a root-mean-square deviation (RMSD) of \SI{2.89}{\milli \hartree} and \SI{2.73}{\milli \hartree}, respectively.
For context, the CCSD(T) exhibits a RMSD of \SI{1.65}{\milli \hartree} for these molecules. The mean absolute deviations (MAD) of \SI{1.39}{\milli \hartree} (CCSD(T)), \SI{1.85}{\milli \hartree} (ph-AFQMC) and \SI{1.84}{\milli \hartree} (ph$^{*}$-AFQMC) show a smaller difference between the methods indicating that some outliers taint the overall performance of AFQMC.
The mean signed deviation (MSD) of \SI{1.39}{\milli \hartree} (CCSD(T)) and \SI{1.17}{\milli \hartree} (ph$^{*}$-AFQMC) indicate under-correlation in contrast to ph-AFQMC with an average over-correlation of \SI{-0.38}{\milli \hartree}.
The modified approach (ph$^{*}$-AFMQC) is also closer to CCSD(T) with an RMSD of \SI{2.66}{\milli \hartree} compared to ph-AFQMC with an RMSD of \SI{3.67}{\milli \hartree}.

Next, we present a more detailed analysis based on the graphical representation of the results in Fig.~\ref{fig:heat}. Let us focus on the more-widely-used CCSD(T) method first. Fig.~\ref{fig:heat} illustrates that CCSD(T) systematically under-correlates compared to the reference values. The most frequent deviation lies around \SI{1}{\kilo \cal / \mol} (i.e. \SI{1.59}{\milli \hartree}). The worst case is the CN molecule with a deviation of \SI{4.1}{\milli \hartree}. The reason for the larger deviation is probably the large spin contamination in the UHF wavefunction.\cite{Bomble2005Heat}

The ph-AFQMC and ph$^{\ast}$-AFQMC results show similar trends. The O$_2$ molecule is the worst case for AFQMC with an undercorrelation of \SI{8.6}{\milli \hartree}. Solving this problem requires a multi-determinant trial wavefunction, for example from a complete active space self-consistent field (CASSCF) calculation in the open-shell subspace. A similar problem in open-shell atoms is solved by using multi-determinant CASSCF trial wavefunctions or by using symmetry restoration techniques.\cite{Motta2018AFQMCREVIEW} Excluding O$_2$ from the statistics reduces the RMSD to 2.41 and \SI{2.11}{\milli \hartree} for the ph-AFQMC and ph$^{*}$-AFQMC method, respectively.

In contrast to the O$_2$ molecule, the ph-AFQMC values for CN, CO, and CO$_2$ molecules show considerable over-correlation (up to \SI{7}{\milli \hartree} for the CN molecule). The origin of these deviations is the fixed-node error. Our modified approach (ph$^{*}$-AFQMC) systematically increases the energies compared to ph-AFQMC. It is noteworthy that the differences are larger in the cases where the ph-AFQMC shows larger over-correlation. The fixed-node errors are therefore significantly reduced for the CN, CO, and CO$_2$ molecules but the residual errors are still sizable. Better trial wavefunctions could further reduce the fixed-node errors.

\begin{figure}
    \centering
    \includegraphics[width=\columnwidth]{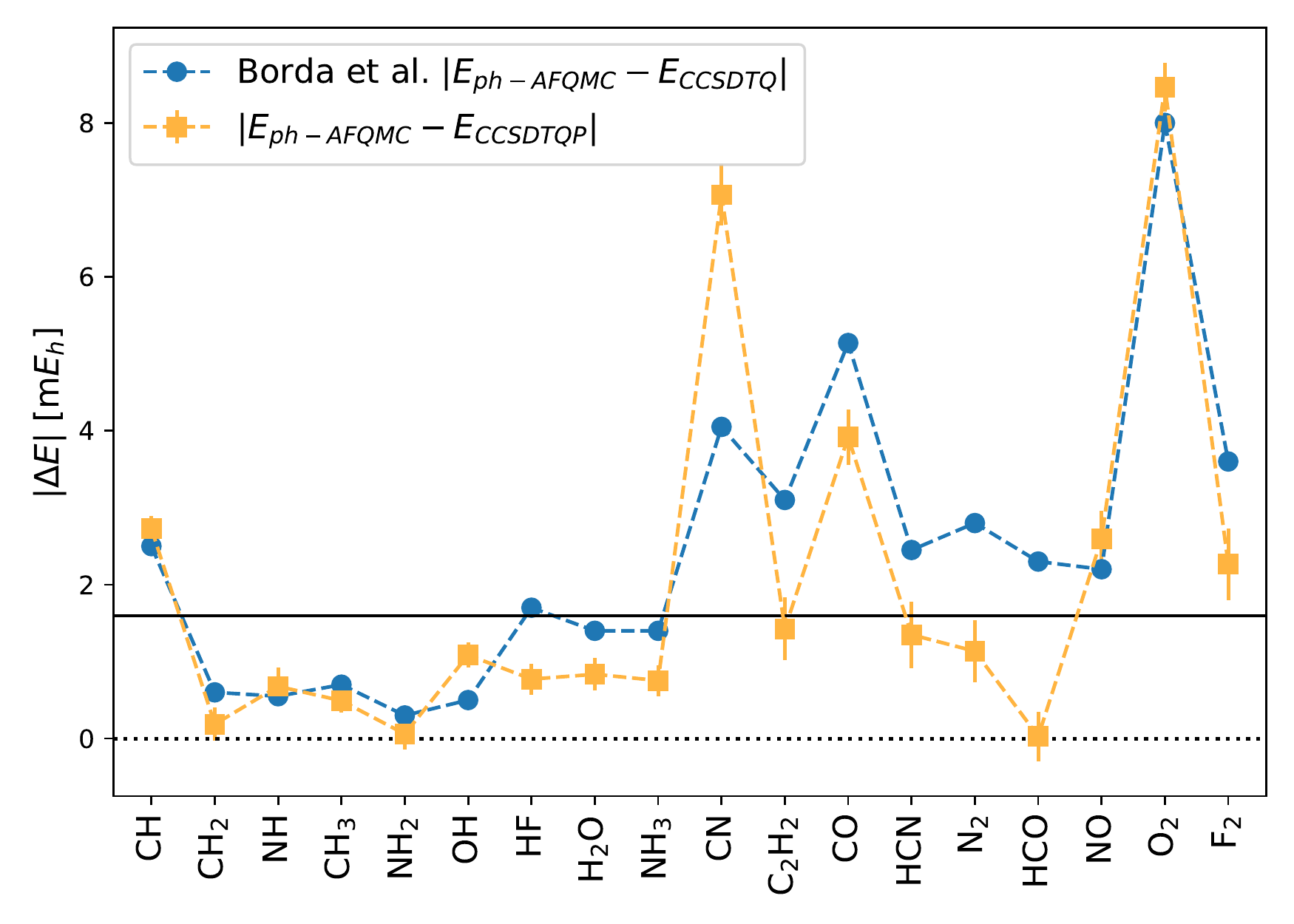}
    \caption{Absolute total energy difference $|\Delta E|$ between ph-AFMQC (QMCPACK) and CCSDTQ (see Ref.~\onlinecite{Borda2019NONSD}) as well as ph-AFQMC (QMCFort) and CCSDTQP for the overlapping molecules in the G1 and HEAT set.}
    \label{fig:morales_comp}
\end{figure}

Borda \emph{et. al.}\cite{Borda2019NONSD} performed similar benchmarks using the G1 test set 
that partially overlaps with the HEAT set. 
They also used cc-pVDZ basis sets with the frozen-core approximation and compared AFQMC total 
energies calculated with the QMCPACK package to the CCSDTQ energies. 
While we used UHF trial wavefunctions, they used ROHF ones for the AFQMC calculations.
Fig.~\ref{fig:morales_comp} shows the deviation of ph-AFQMC from the reference energies.
Overall, most energies agree within 1~m$E_{\rm h}$ between the two ph-AFQMC calculations.
For the remaining cases, our results appear to be closer to the reference result except for the CN molecule.
Possible reasons for these differences include:
(i) smaller statistical and systematic errors in the present work,
(ii) different molecular geometries, 
and (iii) different trial wavefunctions.
The latter matters particularly for the CN molecule because the large spin contamination in the UHF wavefunction makes the ROHF wavefunction a better choice for the trial wavefunction.

\begin{figure*}
    \centering
    \includegraphics[width=0.9\textwidth]{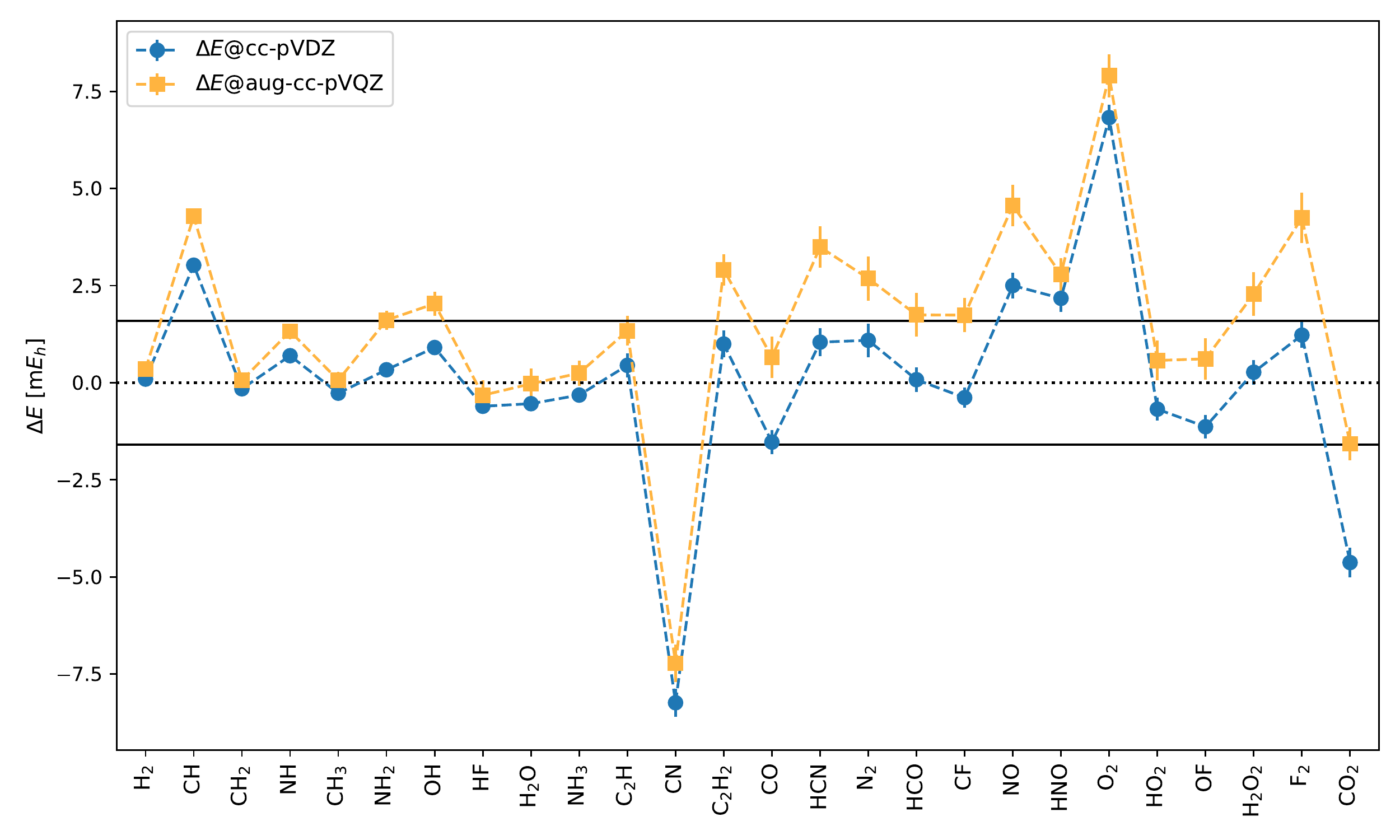}
    \caption{Total-energy difference $\Delta E$ between ph$^{*}$-AFQMC and CCSD(T) for the HEAT set. The cc-pVDZ calculations employ the frozen-core approximation; the aug-cc-pVQZ ones include all electrons. Trial wavefunctions are single RHF and UHF Slater determinants for closed- and open- shell molecules, respectively. The solid black line represents chemical accuracy ($\Delta E < \SI{1}{\kilo \cal / \mol}$).}
    \label{fig:heat_diff}
\end{figure*}

Next, we study the basis-set impact by comparing the total energy differences between ph$^{*}$-AFQMC and CCSD(T) for the cc-pVDZ basis set and the roughly six-times larger aug-cc-pVQZ one. The latter includes core electrons overcoming the frozen-core approximation of the smaller basis. Fig.~\ref{fig:heat_diff} shows remarkably similar energy differences for both basis sets. Three notable exceptions are F$_2$, O$_2$, and OH where the deviation is slightly larger than \SI{1}{\kilo \cal / \mol}. This result is important for two reasons: First, the comparison between ph$^{*}$-AFQMC and CCSD(T) is sufficiently accurate using a cc-pVDZ basis set. Hardly any new conclusion could be drawn from the complete basis-set limit. Secondly, the ph$^{*}$-AFQMC precisely describes core-electron effects at the accuracy of the CCSD(T) method.

\subsection{Benzene Molecule}
\label{BenzeneResults}

\begin{figure}
    \centering
    \includegraphics[width=\columnwidth]{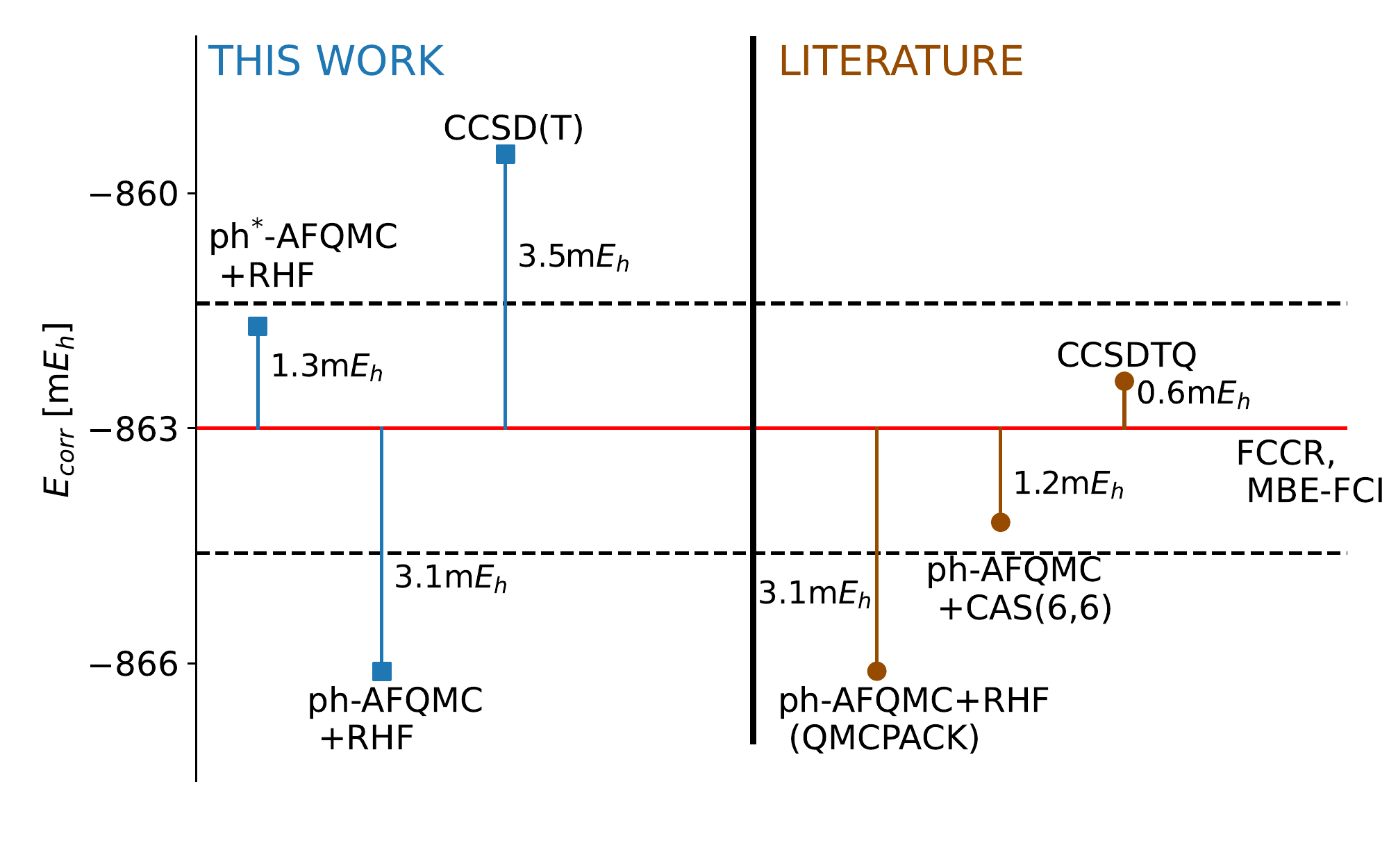}
    \caption{Correlation energies for the benzene molecule using the cc-pVDZ basis set and the frozen-core approximation. 
    ph-AFQMC(RHF), ph$^{*}$-AFQMC(RHF) correlation energies are obtained using QMCFort code, while the CCSD(T) energy is calculated using PySCF code.\cite{pyscf} 
    Other values are taken from the Refs.~\onlinecite{Benzene2020Blindtest,Lee2020BenzeneAFQMC}. The red line represents the reference correlation energy of \SI{-863.0}{\milli \hartree}, while the dashed black lines represent the range of the chemical accuracy (\SI{1}{\kilo \cal / \mol}). 
    All energy values are given in \si{\milli \hartree}.
    }
    \label{fig:benzene}
\end{figure}

The ground state of the benzene molecule with double-zeta basis set (cc-pVDZ) is an interesting system to benchmark state-of-the-art correlation-consistent methods because it is among the largest systems that can be treated directly using FCI diagonalization. 
The benzene molecule in the cc-pVDZ basis set contains 108 orbitals and 30 electrons (in the frozen-core approximation).

Eriksen \emph{et. al.}\cite{Benzene2020Blindtest} performed a blind test on the benzene molecule comparing ten different methods. 
They include the coupled-cluster expansion methods, different selected CI methods, and quantum Monte Carlo methods. 
The authors agreed that the full coupled-cluster reduction (FCCR) and many-body FCI expansion (MBE-FCI) essentially yield the exact correlation energy of \SI{-863.0}{\milli \hartree}.
Lee \emph{et. al.}\cite{Lee2020BenzeneAFQMC} supplemented the study with ph-AFQMC results using two different trial wavefunctions: RHF wavefunction and CASSCF(6,6) multiconfigurational wavefunction.
The ph-AFQMC+RHF overestimates the correlation energy by \SI{3.1}{\milli \hartree}, while the ph-AFQMC+CAS(6,6) over-correlates by \SI{1.2}{\milli \hartree}.
For the sake of completeness, we augmented the study with CCSD(T) results which are under-correlated by \SI{3.5}{\milli \hartree}.
Our ph-AFQMC+RHF result is equivalent to the ph-AFQMC+RHF result calculated using the QMCPACK code.\cite{QMCPACK2020,QMCPACK} 
This serves as an additional validation of our AFQMC implementation in QMCFort.
Similar to CN, CO, and CO$_2$, ph$^{*}$-AFQMC+RHF reduces the absolute value of the correlation energy considerably and leads to an energy under-correlated by \SI{1.3}{\milli \hartree}.
In this case the quality of the ph$^{*}$-AFQMC+RHF is seemingly similar to the quality of ph-AFQMC+CAS(6,6). 
However, better trial wavefunctions promise improved accuracy and, most importantly, more controlled results.
All methods and respective correlation energies are visualized in Fig.~\ref{fig:benzene}. 

\subsection{Water Clusters}
\label{waterclusters}

Great effort has been put into developing empirical models that faithfully represent the properties of bulk water.
It turns out that a good model must also adequately describe the small water clusters present in the Earth's atmosphere. 
Temeslo \emph{et. al.} collected representative structures for water clusters of various sizes.\cite{TemesloWaterClusters2011}
They investigated which contributions are important to obtain accurate formation energies.
Motta \emph{et al.} reported deviations of AFQMC and CCSD(T) larger than the statistical fluctuations.\cite{Motta2109-AFQMC-ED}
Here, we revisit these clusters to scrutinize the accuracy of AFQMC for water.

We estimate the binding energy of the most stable water cluster with
$n \le 5$ H$_2$O molecules 
\begin{equation}
    E_b(n\text{H}_2\text{O}) = E(n\text{H}_2\text{O}) - nE(\text{H}_2\text{O})~.
\end{equation}
Here, $E(n\text{H}_2\text{O})$ is the total energy of the cluster, and 
$E(\text{H}_2\text{O})$ is the ground-state energy of the water molecule.
The cluster geometries are taken from Ref.~\onlinecite{TemesloWaterClusters2011}.
We relaxed the single water molecule using MP2 at aug-cc-pVDZ basis set and obtain a bond length of 0.96593~\AA{} and a bond angle of 103.866$\degree$.
We used the heavy-augmented basis set (aug-cc-pVDZ for O atom and cc-pVDZ for H atom) and all-electron wavefunctions to compare our results with Ref.~\onlinecite{Motta2109-AFQMC-ED}.
The PySCF package\cite{pyscf} produced the CCSD(T) correlation energies and QMCFort the RHF, MP2, and AFQMC ones.
 We truncated the Cholesky decomposition of the ERIs with a threshold of $10^{-6}$\si{\hartree}.
 We propagated 4\,800 walkers for 140\,000 steps with a time-step of \SI{0.01}{\hartree^{-1}}.
 The exchange energy was evaluated after every 100 steps.
 For the largest cluster, we used half as many walkers.
 This setup keeps the systematic errors in the binding energy within \SI{0.05}{\kilo \cal / \mol}.

Table \ref{tab:wclusters} lists RHF, MP2, CCSD(T), ph-AFQMC, ph$^{*}$-AFQMC and AFQMC binding energies from Ref. \onlinecite{Motta2109-AFQMC-ED}.
While the RHF binding energies differ significantly from the correlation-consistent methods, 
all other methods agree within chemical accuracy. 
Our AFQMC values are in better agreement with MP2 and CCSD(T) values than the previous AFQMC values.
The large statistical errors in the reference data might partially explain the discrepancy in AFQMC 
binding energies.
However, the systematic under-correlation of the reference AFQMC binding energies indicates
the existence of systematic errors, too.
One possible source of the systematic errors could be the looser threshold in the Cholesky decomposition of the ERIs ($10^{-4}$\si{\hartree}) in Ref. \onlinecite{Motta2109-AFQMC-ED}.

\begin{table*}
\caption{\label{tab:wclusters} Binding energies of the four most stable water clusters containing up to 5 H$_2$O molecules calculated using heavy-augmented cc-pVDZ basis set and  all-electron wavefunctions. RHF, MP2, CCSD(T), and different ph-AFQMC values in $\text{kcal}/\text{mol}$ are reported.}
\begin{ruledtabular}
\begin{tabular}{l*{6}{d}}
Cluster             & \multicolumn{1}{c}{RHF }& \multicolumn{1}{c}{MP2} & \multicolumn{1}{c}{CCSD(T)} & \multicolumn{1}{c}{ph-AFQMC} & \multicolumn{1}{c}{ph$^{*}$-AFQMC} & \multicolumn{1}{c}{ph-AFQMC\cite{Motta2109-AFQMC-ED}} \\ \hline
2Cs                 & -3.82    & -5.22    & -5.18   & -5.17(5)    & -5.06(7)    &  -5.11(31)        \\
3UUD                & -10.52   & -15.83   & -15.62  & -15.67(9)   & -15.68(9)   &  -14.78(64)       \\
4S4                 & -19.00   & -28.36   & -27.87  & -28.12(10)  & -28.11(10)  &  -26.49(46)       \\
5CYC                & -25.30   & -37.48   & -36.78  & -37.14(28)  & -37.31(28)  &  -36.27(59)       \\ 
\end{tabular}
\end{ruledtabular}
\end{table*}

To test ph-AFQMC and ph$^{*}$-AFQMC for size consistency we dissociated the cluster into individual molecules by shifting the second and third molecule in the 3UUD cluster by 8~\AA{} in $x$ and $y$ direction, respectively.
We computed the total energy of the stretched cluster and of the individual water molecules individually.
We summarize our findings in Table~\ref{tab:3UUDdis}.
Within statistical fluctuations, all fragments exhibit the same total energy which is equal to a third of the stretched cluster.
This demonstrates that ph-AFQMC and ph$^{*}$-AFQMC are size-consistent.

\begin{table}
\caption{\label{tab:3UUDdis}
The AFQMC energies of the stretched 3UUD cluster and its fragments, and the binding energies demonstrate that ph-AFQMC and ph$^{*}$-AFQMC are both size-consistent (all values are given in \si{\hartree}).}
\begin{ruledtabular}
\begin{tabular}{l*{2}{d}}
Cluster             & \multicolumn{1}{c}{ph-AFQMC} & \multicolumn{1}{c}{ph$^{*}$-AFQMC}             \\ \hline
3UUD                & -228.82750(12)    & -228.82448(12)      \\
1-H$_2$O            &  -76.27576(4)     &  -76.27475(4)       \\
2-H$_2$O            &  -76.27580(4)     &  -76.27477(4)       \\
3-H$_2$O            &  -76.27585(4)     &  -76.27481(4)       \\ \hline
$E_b$               &    0.00009(14)    &    0.00015(14)
\end{tabular}
\end{ruledtabular}
\end{table}

\section{Conclusion}
\label{conclusion}

The phaseless auxiliary-field quantum Monte Carlo is increasingly popular due to its high accuracy, the low polynomial scaling ($N^{3} - N^{4}$), and its applicability to quantum chemistry and condensed matter physics. Using the DFT or HF solutions as starting point, it can be considered as a natural extension of these methods with similar scaling but higher accuracy. 

We have presented a Fortran implementation of the AFQMC, QMCFort, that enables efficient large-scale calculations on CPUs. 
The code is parallelized using MPI, OpenMP, and parallel BLAS, and runs near peak performance for typical systems (see Fig.~\ref{fig:perf}). 
QMCFort can run AFQMC simulations independently or obtain Cholesky vectors of the two-electron intermediates via interfaces to VASP and PySCF. 

Using QMCFort, we compared the accuracy of the ph-AFQMC method to the "gold-standard" CCSD(T) and the more accurate CCSDTQP method. 
For this purpose, we calculated the ph-AFQMC energies of the 26 molecules in the HEAT set, the benzene molecule, and water clusters.
For the HEAT set, the ph-AFQMC yields a mean absolute deviation (MAD) of \SI{1.85}{\milli \hartree} (\SI{1.15}{\kilo \cal / \mol}) compared to CCSDTQP values. 
The poor performance for four molecules (CN, CO, CO$_2$, O$_2$) is responsible for a large part of this deviation. Such outliers are clearly troublesome as they are difficult to recognize in the absence of reference calculations. 
Furthermore, CCSD(T) is certainly more robust  for the HEAT set and is highly accurate even if the  groundstate wavefunction involves significant double excitations. 
One possible explanation for the presence of outliers and the failure of the AFQMC is that the method fails to account accurately for strong double excitations.
This suggests that the AFQMC with a single determinantal trial wavefunction is only capable to treat weakly correlated materials that lack strong static correlation effects even in the space of double excitations.

For the remaining molecules in the HEAT set, ph-AFQMC performs similarly to or better than the CCSD(T). 
We modified the phaseless approximation---ph$^{*}$-AFQMC---to overcome at least partly the over-correlation problems often encountered with ph-AFQMC. 
For the CN, CO, and CO$_2$ molecules, where the over-correlation effects are particularly pronounced, the modified approach indeed improves the energies significantly.
In the case of the benzene molecule, ph$^{*}$-AFQMC yields a correlation energy \SI{1.3}{\milli \hartree} higher than the FCI reference value. 
This is noticeably better than ph-AFQMC and comparable to the accuracy of ph-AFQMC with a multi-determinant CAS(6,6) trial wavefunction. Finally,
the AFQMC binding energies of water clusters agree with the CCSD(T) and MP2 binding energies to within \SI{0.5}{\kilo \cal / \mol}. For the four water clusters, our present results are generally closer to CCSD(T) results than previous AFQMC results, substantiating our claim that the AFQMC method with a single Slater determinant is particularly suitable for weakly correlated molecules and materials.

Both the ph$^{*}$-AFQMC, as well as the ph-AFQMC, are rather ad hoc approaches to deal with the fermionic sign problem and the exponential increase of the noise.
We feel that the present work does not conclusively show that the ph$^{*}$ approximation is superior to the ph approximation.
However, the present work clearly shows that even minor changes to the phaseless approximation can affect the final results.
Although we could not find an approximation that more generally improves the results, we believe that further research in this direction is well warranted.

In summary, the ph-AFQMC is a promising and reliable method with high accuracy and reasonable computational cost.
It provides nearly chemical accuracy for the small molecules studied in this work if the molecules are weakly correlated.
We plan to further improve the accuracy of \mbox{QMCFort} through algorithmic improvements and the use of more accurate trial wavefunctions.

\section{Acknowledgments}
Funding by the Austrian Science Foundation (FWF) within the project P 33440 is gratefully acknowledged. All calculations were performed on the VSC4 / VSC5 (Vienna scientific cluster). 

\section*{Author declarations}
\subsection*{Conflict of Interest}
The authors have no conflicts to disclose.

\section*{Data availability}
The data that support the findings of this study are available within the article.

\appendix* 

\section{Interface with VASP}

Consider the set of mean-field orbitals $\{ \phi_{p}(\textbf{r}) \}$.
The single-particle Hamiltonian matrix elements are computed as
\begin{equation}
    h_{pq} = \int \text{d}^3(\textbf{r}) \; \phi_{p}^{*}(\textbf{r}) \hat H_1 \phi_{q}(\textbf{r})~,
\end{equation}
where $\hat H_1$ contains the kinetic energy and the Coulomb attraction between an electron and the nuclei.
To obtain objects equivalent to Cholesky vectors $L_{g,pq}$, we introduce two-orbital densities 
\begin{equation}
    \rho_{\textbf{G},pq} = \frac{1}{\sqrt{\Omega}} \int \text{d}^{3}(\textbf{r}) \; e^{i\textbf{G}\textbf{r}} \; \phi_{p}^{*}(\textbf{r}) \phi_{q}(\textbf{r})~,
\end{equation}
where $\{ \textbf{G} \}$ is the set of reciprocal lattice vectors defined by the cutoff energy $\textbf{G}^2/2 \leq E_{\text{cut}}$ and $\Omega$ is the volume of the system. The relation between Cholesky vectors and two-orbital densities is
\begin{equation}
    \sum_g^{N_g} L_{g,pq} L_{g,rs} = \sum_{\textbf{G}}^{N_{\textbf{G}}} \frac{4\pi}{\textbf{G}^2} \; \rho_{\textbf{G},pq} \rho_{\textbf{G},rs}~,
\end{equation}
where $N_g \leq \text{min}(N_{\textbf{G}}, N^2)$. 
In Ref.~\onlinecite{AmirAFQMC}, we introduced a truncated singular value decomposition to obtain the Cholesky vectors
\begin{equation}
    L_{g,pq} = \text{SVD} \left( \frac{\sqrt{4\pi}}{|\textbf{G}|} \rho_{\textbf{G},pq} \right)~.
\end{equation}

\bibliographystyle{aapmrev4-1}
\bibliography{bibliography}         

\end{document}